\documentclass[]{spie}  

\usepackage{amsmath,amsfonts,amssymb}
\usepackage{graphicx}
\usepackage[colorlinks=true, allcolors=blue]{hyperref}

\title{Lynx grating spectrometer design: Optimizing chirped transmission gratings}

\author[a]{Hans Moritz G\"unther}
\author[a,b]{Ralf K. Heilmann}
\affil[a]{MIT Kavli Institute for Astrophysics and Space Research, Massachusetts Institute of Technology, Cambridge, MA 02139, USA}
\affil[b]{Space Nanotechnology Laboratory, Massachusetts Institute of Technology, Cambridge, MA 02139, USA}

\authorinfo{Send correspondence to H.M.G. (E-mail: hgunther@mit.edu)}

\begin{document} 
\maketitle

\begin{abstract}
Lynx is one of four large-mission concept studies for NASA's 2020 Decadal survey. The design reference mission includes an X-ray grating spectrometer (XGS) based on critical-angle transmission (CAT) gratings. In the past we studied different grating sizes and arrangements using traditional flat CAT gratings with constant bar spacing. However, new technology development brings chirped gratings in reach. Using chirped gratings where the grating bar spacing varies over a grating allows us to fill the aperture with larger gratings because the chirp can compensate for some aberrations caused by the deviation of large flat gratings from the Rowland torus. This reduces the area blocked by grating support structures. Using larger gratings also carries potential cost savings.
We use ray-tracing to study an XGS design with chirped grating and find that using chirped gratings of $80 * 160$~mm size allows us to reduce the number of gratings from a few thousand to a few hundred, while simultaneously increasing the effective area by 25\% and keeping the resolving power constant. Bending those gratings to maintain a constant blaze angle over the entire grating increases the effective area by another 5-10\%.

\end{abstract}


\section{INTRODUCTION}
\label{sec:intro}
High resolution X-ray spectroscopy is a well-established method to study a wide variety of phenomena in the high-energy universe from early phases of star formation to the outflows of massive black holes in the center of far away galaxies. Often, high-resolution X-ray spectroscopy reveals information than cannot be obtained by any other method. For example, the density of the accretion shock in young stars can only be measured in He-like triplets of O~{\sc vii} and Ne~{\sc ix}, located around 21 and 13~\AA{}, respectively. Despite great advances in X-ray microcalorimeters, only diffraction gratings can deliver a resolving power $> 1000$ across the soft X-ray band.

In preparation for the 2020 Decadal survey, NASA organized a detailed study of four Surveyor-type missions. One of them focused on the X-ray band, and is called ``Lynx''\cite{gaskin}. The design
reference mission (DRM) for Lynx includes a mirror with a 2~m$^2$ collecting area at
1~keV and a point-spread-function (PSF) of 0.5~arcsec half-power-diameter
(HPD). Lynx would have two instruments at the focal point with different field-of-view and energy resolution (High-Definition X-ray Imager
(HDXI)\cite{HDXI} and a microcalorimeter\cite{MICROCAL}) and retractable gratings that can be inserted into the beam to diffract photons to a separate detector (X-ray grating spectrometer - XGS). While reflection gratings have been considered\cite{OPXGS}, they have not yet demonstrated sufficient resolving power\cite{2020ApJ...897...92D}, and the DRM features critical-angle transmission (CAT) gratings\cite{Heilmann:19}.

The design, hardware requirements, and predicted performance of the XGS in the DRM are described in detail in Ref.~\citenum{CATXGS}; in the same reference we also explain the setup of our ray-trace simulations. A more detailed treatment of diffraction by the grating support structures was added in Ref.~\citenum{10.1117/12.2525814}. In section~\ref{sect:raytrace} we present a short summary of the setup, but refer the reader to Ref.~\citenum{CATXGS} and Ref.~\citenum{10.1117/12.2525814} for more details. 

The goal of this paper is to study a scenario in detail that was only briefly mentioned in Ref.~\citenum{CATXGS}: CAT gratings with a ``chirp'', i.e.\ a spatially variable grating period $d$. We discuss the motivation to use chirped gratings in section~\ref{sect:motivation} before we go into results for ray-traces with chirped gratings (section~\ref{sect:chirp}). Additionally, one can combine a chirp with physically bending the gratings, and we demonstrate in section~\ref{sect:bend} that this increases the effective area by another 5-10\%. We end with a discussion and summary in section~\ref{sect:summary}.

\section{SETUP FOR RAY-TRACES}
\label{sect:raytrace}
Our simulations are based on a geometric ray-trace, which follows
individual rays through the system from the entrance aperture to the
detector. Simulations are performed with MARXS
1.2\cite{marxs,marxs1.2}, which is written in Python and licensed
under the GPL v3. It is available on
github\footnote{\url{https://github.com/chandra-marx/marxs}}. Code specific to the Lynx XGS and the analysis shown in this article is also available\footnote{\url{https://github.com/hamogu/marxs-lynx}}; we
used the version with commit hash adb62ab. 

The setup is described in detail in Ref.~\citenum{CATXGS}. In addition, improvements on the treatment of the grating support structures, which act as diffraction gratings themselves, are described in Ref.~\citenum{10.1117/12.2525814}. For the analysis here, we assume that the parameters of the grating bar support structures do not depend on the size of the grating, i.e.\ that larger gratings do not require any thicker support structures than we assumed for the $50*50$~mm$^2$ gratings discussed in Ref.~\citenum{CATXGS}. In Ref.~\cite{CATXGS} we investigated different options for sub-aperturing and found that covering 2/3 of the aperture with a mix of $20*50$~mm$^2$ and $50*50$~mm$^2$ gratings is sufficient to reach the Lynx science requirement at 0.6~keV for the effective area of the XGS $A_{\mathrm{eff}} > 4000$~cm$^2$ with margin. As the same time, sub-aperturing increases resolving power $R$ compared to a scenario where the entire aperture is filled with gratings from about 6000 to 8000. Yet, Lynx is still in an early stage of development and changes to mirror design or science requirements might leads us to cover more or less of the aperture with gratings. For the study presented here, we want to avoid the complication of presenting every plot with different options for the sub-aperturing angle. Thus, we simply use a full aperture as our baseline and study the impact of chirping and bending gratings relative to that. That means that the numbers for $R$ and $A_{\mathrm{eff}}$ shown for the baseline match the ``full-aperture'' scenario from Ref.~\citenum{CATXGS}, not the exact numbers from the DRM.

The CAT gratings for the Lynx XGS are developed at the
MIT Space Nanotechnology Laboratory
\cite{Heilmann:11,doi:10.1117/12.2188525,10.1117/12.2314180,10.1117/12.2529354}. The high aspect-ratio grating bars are assumed to be
5.7~$\mu$m deep and supported by an L1 support structure running
perpendicular to the grating bars themselves, and the entire membrane
(bars and L1) is mechanically stabilized by a hexagonal L2 support
structure, which is 0.5~mm deep. 
Preliminary structural analysis indicates that changes in the shape of the L1 and L2 support structures could be used to reduce the thickness and covering fraction of the support structures compared to the current design, so our assumption that the L1 bars and L2 hexagons currently planned work for larger gratings as well seems reasonable for this early stage of the spectrograph design.
Absorption and diffraction of photons
by the L1 and L2 support structures is included in our
simulations. The grating membrane is surrounded by a narrow solid Si frame. We also explore if bending the gratings into the shape of a cylinder (with the axis of the cylinder parallel to the grating bars) can further improve performance. Smaller gratings with bars 4.0~$\mu$m deep and a size of $10 * 30$~mm$^2$ have been bent in the laboratory and were found to maintain grating efficiency with no obvious mechanical damage\cite{10.1117/12.2274205}.

\section{Motivation for chirped gratings}
\label{sect:motivation}
In the XGS, the diffraction gratings need to be arranged on the surface of a torus, the so-called Rowland torus\cite{beuermann:78}. The Rowland torus is curved, but transmission gratings are usually flat. Thus, for any positioning of gratings, parts of the grating will be above and others below the Rowland torus surface. If a ray intersects with a grating too early, the ray will be diffracted at a larger distance from the focal plane and thus will land further away from the zeroth order, while rays intersecting a grating below the surface of the Rowland torus are diffracted not as far. This effect spreads out the photons and reduces the spectral resolving power. In earlier X-ray instruments with transmission gratings, e.g.\ the Chandra/HETG, this is not a limiting effect. Gratings are small (about 20~mm or so on the side) and the resolving power $R$ requirements are relatively modest compared with the Lynx requirement of $R> 5000$. However, for the Lynx X-ray grating spectrograph (XGS) this is actually an effect that can limit resolving power and drives the choice of grating size in the baseline design\cite{CATXGS}.

Different approaches can be used to compensate the loss in resolving power. The simplest idea might be to just use smaller gratings. Gratings of $20*50$~mm$^2$ are small enough to match Lynx resolving power requirements\cite{CATXGS}. However, several thousand gratings are needed to cover the aperture in this case. A second idea is to bend the gratings so that they are not flat and follow the torus better. So, in principle curved gratings can be used to match the grating position better to the Rowland torus. This is complicated by the fact that we also try to maintain a narrow range of blaze angles to optimize the number of photons diffracted into those grating orders that are covered by detectors. The radius of the Rowland circle is about half of the focal distance. To maintain the ideal blaze angle for all rays in a converging beam, the gratings would have to be bent with a radius close to the focal length, while the radius to match the surface of the torus is half of that.  In addition, the developed CAT grating manufacturing process ideally produces grating bars that are perpendicular to the surface of the grating. In order to maintain a blaze angle around 1.6~degrees, gratings therefore have to be mounted at an angle. These competing requirements means that bent gratings with constant grating bar spacing cannot be used to compensate the loss in $R$ that occurs for larger gratings, unless a significant drop in effective area is accepted.

Another approach is to use flat gratings, oriented to make the blaze condition work at least for the center ray, and to change the grating period $d$ smoothly over the grating ("chirp") such that the part of the grating that is located "below" the surface of the Rowland torus, where photons would be diffracted not far enough, has a slightly smaller $d$ and thus diffract to larger angles. Conversely, the part of the grating located "above" the Rowland torus needs to have a slightly larger $d$. The grating equation gives the relation between $\lambda$, the wavelength of the photon, the grating order $n$, the angle of incidence $\alpha$, and the angle of diffraction $\theta_n$:
$$
n\lambda = d (\sin \theta_n - \sin \alpha ) \; .
$$

\section{Chirped gratings in the XGS}
\label{sect:chirp}
\subsection{General setup}
The DRM for the Lynx XGS uses gratings of $20 * 50$~mm$^2$, where the long direction is perpendicular to the dispersion and is chosen to minimize technical risk, as this size is close to the size of gratings that have been manufactured in the past. The short side is along the dispersion direction and is limited by the aberration discussed above. At this size, over 5000 gratings are required to cover the aperture. For most of the aperture, larger gratings of $50*50$~mm$^2$ are also sufficient\cite{CATXGS}. So, with the added complexity of having gratings of two different sizes, the number of gratings can be brought down to about 3000. In contrast, we here investigate much larger gratings of $60 * 180$~mm$^2$. At this size, only about 500 gratings are required to cover the full aperture. The choice of $60 * 180$~mm$^2$ is driven by the size of commercially available 200~mm wafers, such that one grating fits onto a wafer with comfortable margin for handling and processing.

In the following simulations, we assume a continuous detector with no chip gaps. This is just a computational simplification, such that we can run simulations for just three wavelength points without worrying if any one of them falls into a chip gap.
Ray-racing is run with $10^5$ rays per simulation.

\subsection{Approach to find the chirp}
We chose a reference energy near the middle of the XGS bandpass (0.6~keV). 
For each grating, we calculate the position on the detector, where a ray that passes through the center of the grating (which is placed exactly on the Rowland torus) intersects with the detector. For all other positions on the grating, we can then determine which diffraction angle $\theta$ is required to bring rays to the same position on the detector. From the grating equation, we can determine the value of $d$ required at that grating position. The 3D geometry of this is a little cumbersome, so for simplicity we optimize $d$ numerically instead of tracing the geometry analytically. Numerically, we just perform a ray-trace varying $d$. Since we have a code to run ray-traces anyway, this is simple and fast to implement.

Of course, the diffraction of light requires a periodic structure and would not work if gratings did not have a regular period. However, the required chirp is so small that the period $d$ changes noticeably only over macroscopic scales, thus narrow diffraction orders still occur. This is somewhat analogous to stacking a telescope with a large number of smaller gratings. The grating bars are periodic within each grating, but the gratings are not phased up, and diffraction still happens, as evidenced by the Chandra/HETGS and other observatories.

In principle, the chirp in $d$ is not the same for all wavelength and all angles, but since CAT gratings diffract photons into a relatively narrow range of angles, that is not a problem in practice, as we show below.

\begin{figure} [ht]
\begin{center}
\includegraphics[width=0.7\textwidth]{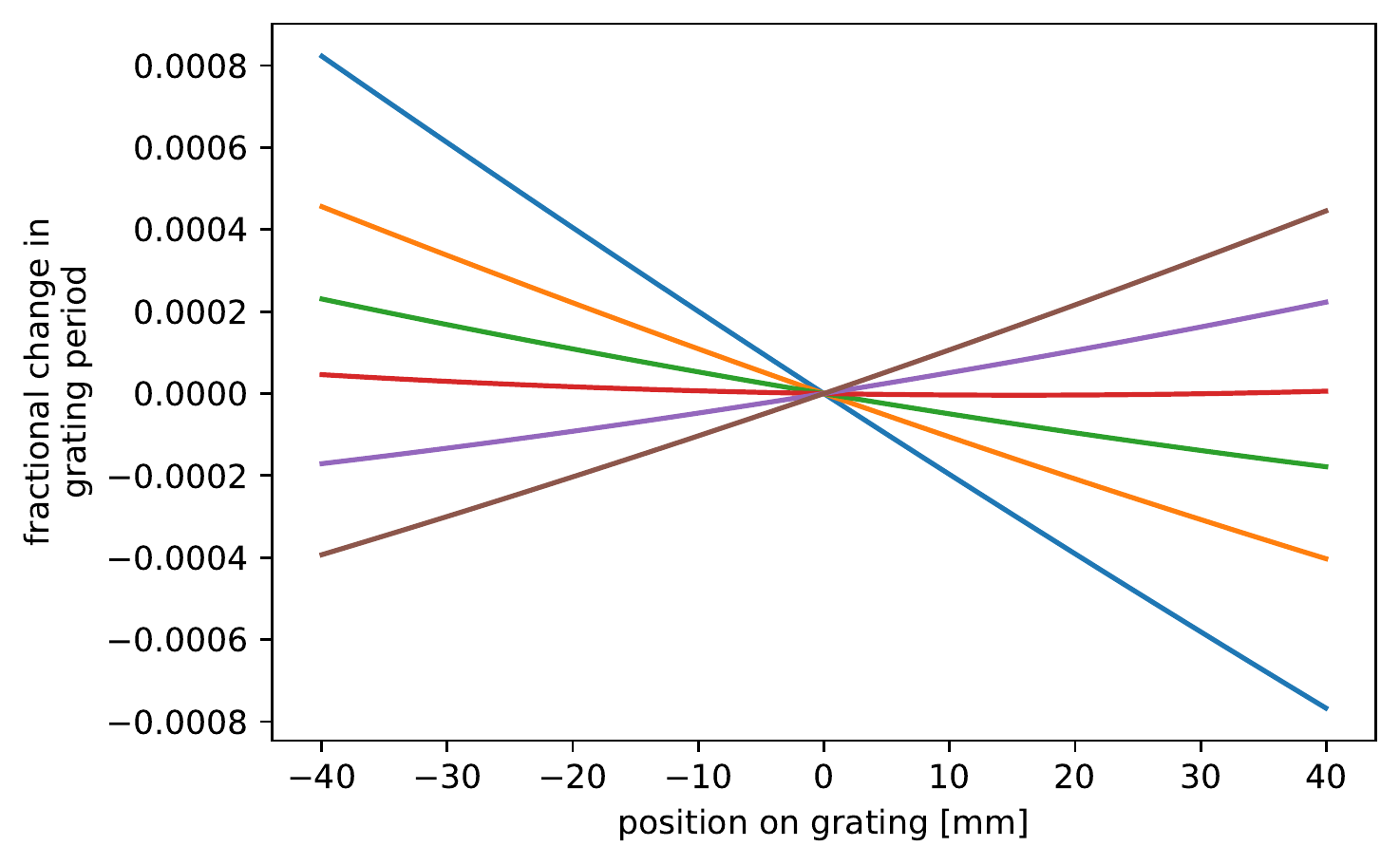}
\end{center}
\caption {\label{fig:chirps}
Relative change in the grating period (chirp), that is required shown for a sample of XGS CAT gratings. The grating period is exactly at the reference period in the center of the grating, thus all lines cross through (0,0). The chirps are calculated numerically for 50 evenly spaced points on each grating.
}
\end{figure}

Figure~\ref{fig:chirps} shows the relative change in the grating period for different gratings as calculated numerically. The gratings are located at different locations on the aperture. Some of them need a decreasing grating period in the direction of the positive $y$-axis, others require an increasing period. What they have in common though, is that all solutions are very close to linear. Thus, from now on, we will impose a linear relationship between the position on the grating and the fractional change in period. This speeds up the calculations significantly, because only a single point is needed to define a line that goes through 0 at the center of the gratings. Beyond numerical convenience, this also simplifies the specification of actual physical gratings. While a different gradient in grating period $d$ might be needed for different gratings, all chirps have a linear relationship.

\subsection{How many different types of gratings do we need?}

The ideal chirp is different for almost every grating. However, from a cost and schedule perspective, it is much preferable to use only few different types of gratings. We can manufacture gratings with a few different chirps and then for each position pick a chirp that is very close to the ideal number. This deviation will lead to some loss of resolving power, but this is acceptable to a certain degree. While the simulations here are done with a perfectly aligned model, in practice, mechanical misalignments, aberrations, and pointing errors all reduce the theoretically achievable resolving power. A small contribution from using gratings with non-ideal chirps does not impact the final resolving power significantly.

\begin{figure} [ht]
\begin{center}
\includegraphics[width=0.7\textwidth]{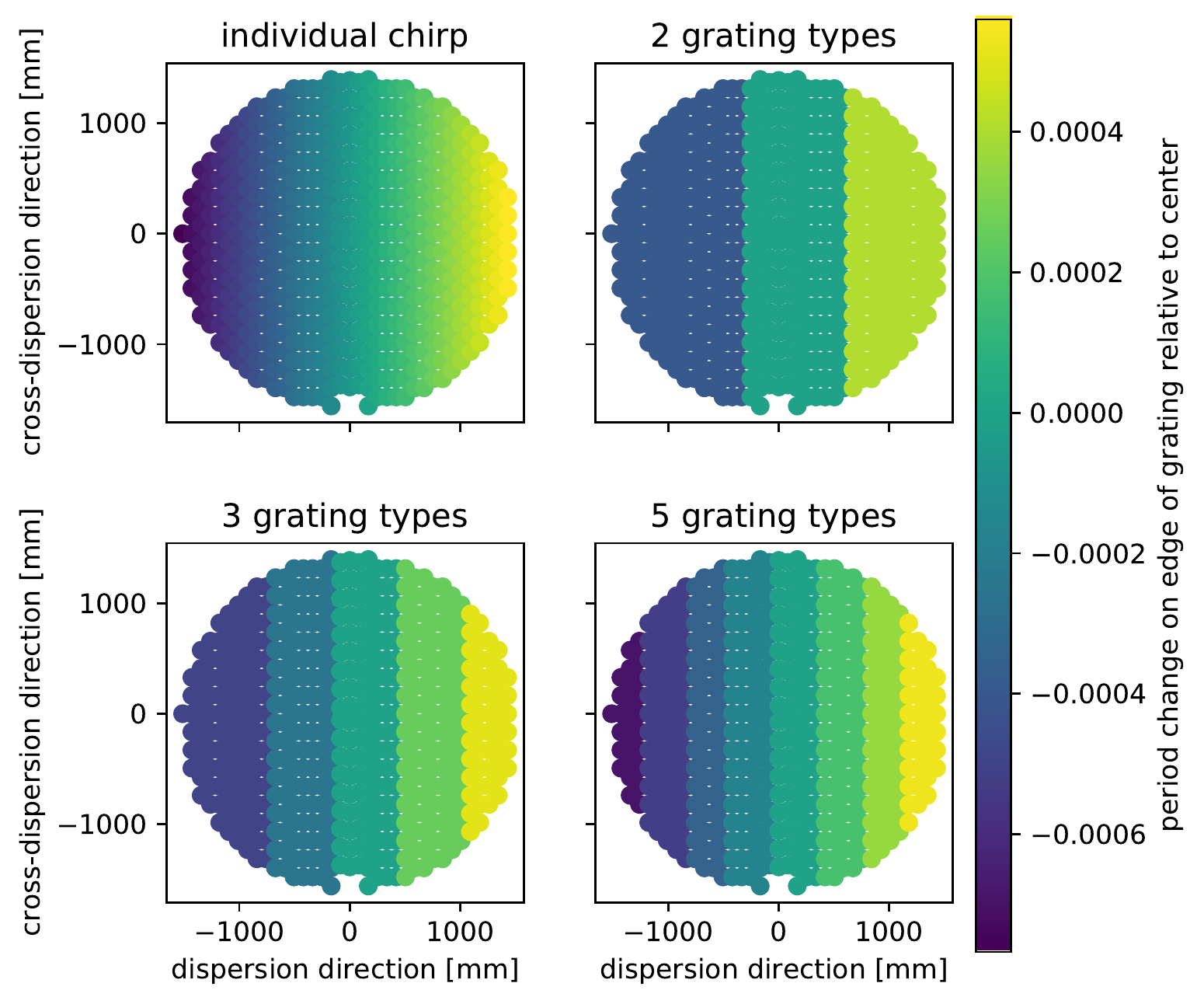}
\end{center}
\caption {\label{fig:chirps_distribution}
This figure shows the distribution of chirps for different simulations. \emph{top left:} Gratings have individual chirps. Note that the required chirp is not symmetric for gratings on the left and right of the figure. \emph{other panels:} Setups with a limited number of grating types where for each position a chirp is chosen that is as close to the ideal chirp as possible. Assuming gratings can be reversed, two different chirps ($0, c$) allow for three different values ($-c, 0, c$) in the figure, thus the top right panel (two grating types) has three stripes and the bottom left panel has five stripes. In the bottom right panel, there are four grating types to the left of the center, but only three to the right, because the chirp goes down to -0.0007, but only to +0.0005 on the other end.}
\end{figure}

In this section, we test how many different chirps are needed. We assume that there will always be  a group of gratings with no chirp. For example, a simulation might use gratings with just three different chirps: -0.0004, 0, 0.0004. Here the value of the chirp is expressed as the relative change of the grating period from the center to one edge. The relative change in $d$ for chirps of -0.0004 and +0.0004 is the same, just in one case it is increasing from left to right, in the other case, it is decreasing. CAT gratings do not have a preferred direction and the same type of grating can be used, just mounted with a different rotation, such that three values of the chirp can be realized with just two different types of gratings (or, e.g., five values of the chirp with just three types of grating, one with constant $d$ and two with different chirps). In practice, grating facets will have a handling layer, screw holes for mounting or similar, so that this has to be taken into account in the mechanical design, but this is still simpler than manufacturing, testing, etc.\ more different grating types.
Figure~\ref{fig:chirps_distribution} shows the ideal chirp (top left) for all gratings in the aperture, as well as the distribution of chirps for arrangements that use 2, 3, or 5 different chirps. The required chirp depends only on the position of the grating in the dispersion direction, so the arrangement forms bands in the image. 

\begin{figure} [ht]
\begin{center}
\includegraphics[width=0.7\textwidth]{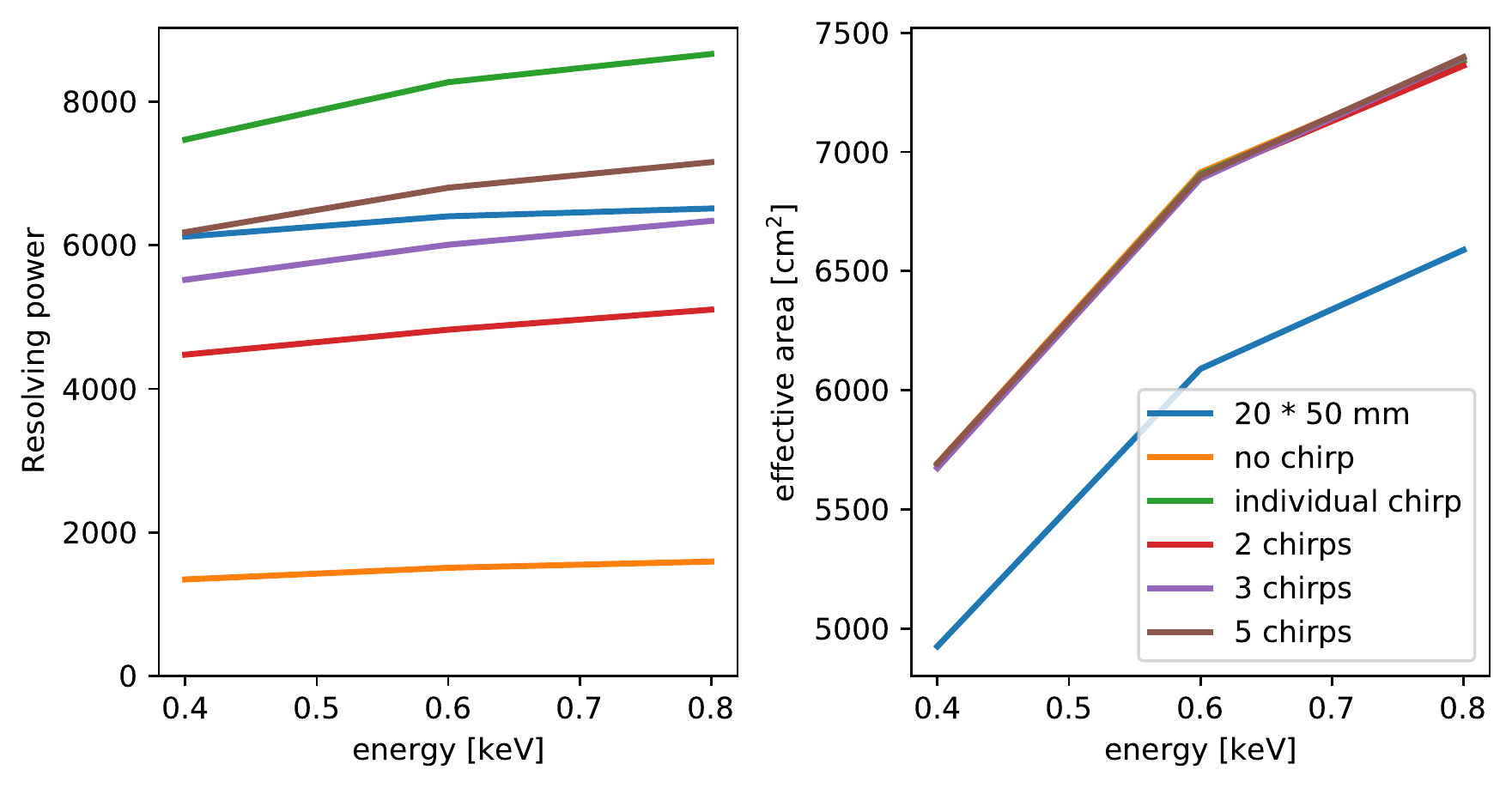}
\end{center}
\caption {\label{fig:chirp_aeff_res}
Average resolving power $R$ (left) and effective area $A_{\mathrm{eff}}$(right) for different grating arrangements. Simulations are performed for three energies within the XGS bandpass. 
}
\end{figure}

Figure~\ref{fig:chirp_aeff_res} shows the resolving power $R$ and effective area $A_{\mathrm{eff}}$ for those four scenarios, as well as the baseline with smaller gratings and, for comparison, $160 * 80$~mm$^2$ gratings with no chirp. 
The right plot shows that the effective area increases by about 20\% when larger gratings are used. The exact number depends on the distribution of the gratings over the aperture, the position of the mirror support spider, the size of the grating holders, the mounting structure etc., which can be optimized at a later stage, but it is clear even from this simple model that larger gratings offer a better $A_{\mathrm{eff}}$, while simultaneously being more cost-effective because far fewer elements need to be manufactured, tested, and installed. The increased $A_{\mathrm{eff}}$ can either be used to improve the science output, or, for constant $A_{\mathrm{eff}}$ compared to the baseline design, we could subaperture, increasing $R$ over the number shown in the figure and reducing the number of gratings needed even further.

In all cases, the chirp is calculated for photons of 0.6~keV, but for the angles and diffraction orders relevant for the XGS, the figure shows that the chirp is equally effective for all relevant energies.

\section{Should chirped gratings be bent?}

\begin{figure} [ht]
\begin{center}
\includegraphics[width=0.5\textwidth]{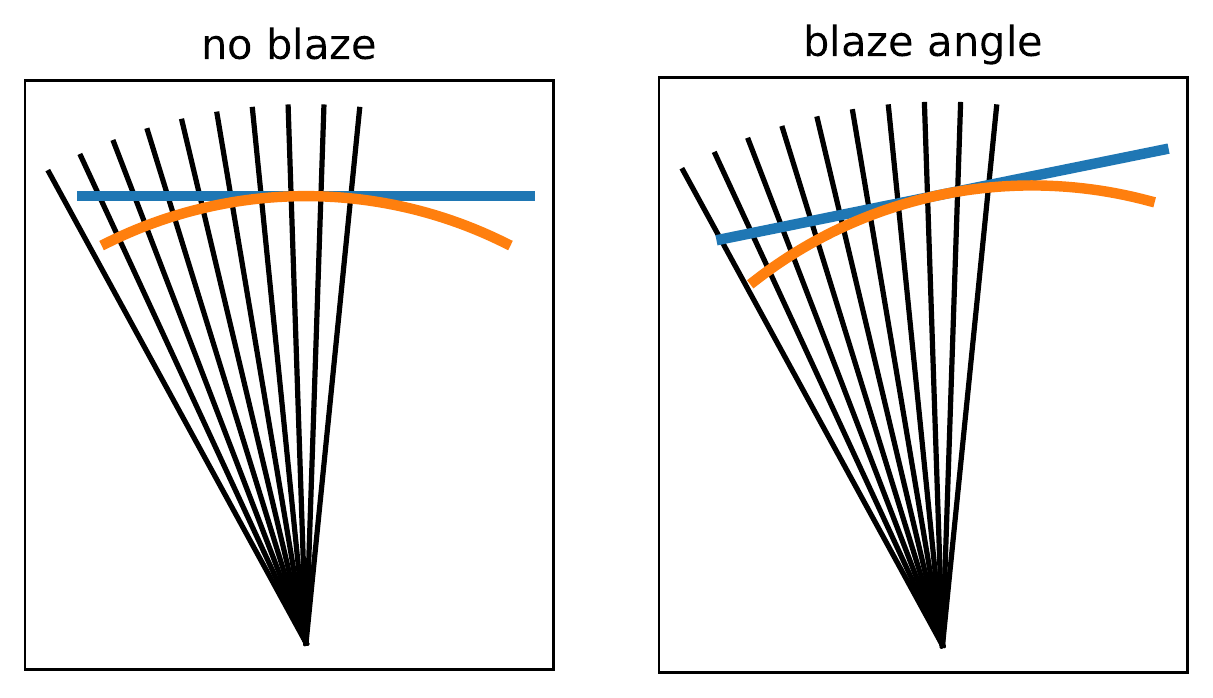}
\end{center}
\caption {\label{fig:explainbending}
In a converging beam (black rays) the angle between the grating normal and the rays changes with grating position for a flat grating (blue). Bending the grating can compensate for this problem (orange) for gratings that are small compared to the radius. This is true for gratings where the ray through the grating center is normal to the grating (left) and for gratings with a design blaze angle (right).
}
\end{figure}

\begin{figure} [ht]
\begin{center}
\includegraphics[width=0.7\textwidth]{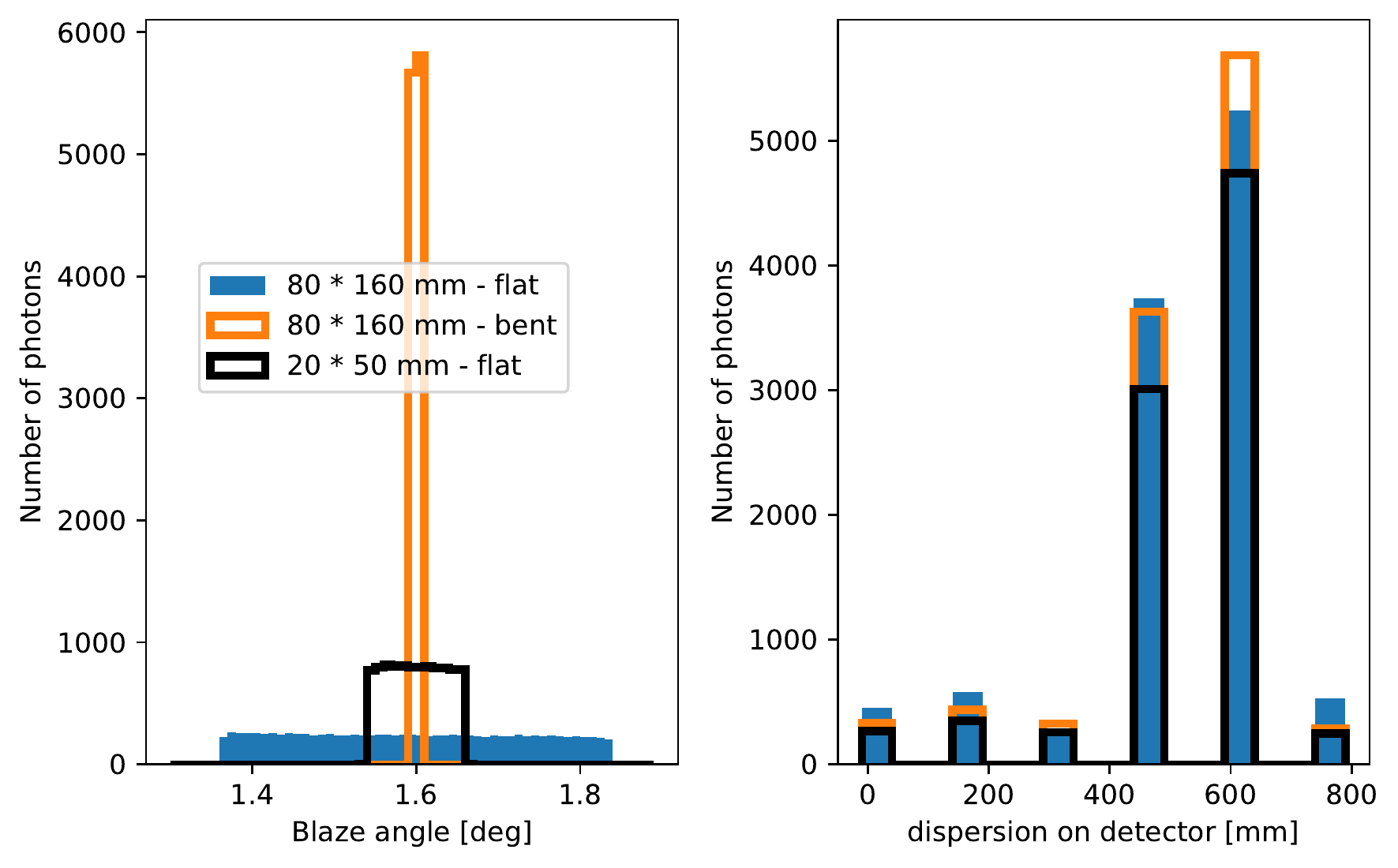}
\end{center}
\caption {\label{fig:blazebend}
\emph{left:} Distribution of blaze angles with small flat, large flat, and large bent gratings. Because of the converging beam, a range of blaze angles is unavoidable for flat gratings - the larger the grating, the larger the range. Bending the grating can compensate this effect. For simplicity of the assembly, we use the same bending radius for all gratings. However, the gratings are arranged on the surface of the Rowland torus and have different distances to the focal point, thus even in the case with bent gratings the blaze angles show a small spread.
\emph{right:} Distribution of photons on the detector for a 0.4~keV simulation. Individual orders are apparent. The zeroth order (direct light) is at detector position 0~mm. The brightest order is order 4 around 600~mm, which has noticeably more signal in the simulation with bent gratings, while flat gratings distribute more photons into other orders.
}
\end{figure}

The probability to diffract a photon into a specific diffraction order depends on the blaze (for CAT gratings, the most likely diffraction angle is about twice the blaze angle). Because rays passing through a large flat grating in a converging beam cover a range of blaze angles (Figure~\ref{fig:explainbending}), the photons are distributed over more diffraction orders than in the case of bent gratings which maintain the design blaze angle over the entire surface of the grating (figure~\ref{fig:blazebend}). The Lynx XGS is designed with 16 detector chips that cover the dispersion coordinate from about 400 to 700 mm. For photons of 0.4~keV that includes just the two most prominent orders. Bent gratings concentrate more light into these orders, and thus raise the effective area of the instrument compared to flat gratings. This is an energy dependent effect, which can increase the effective area by about 5-10\% (see an example in the right of figure~\ref{fig:blazebend}). This is only relevant for large gratings, because on smaller gratings the rays have a smaller range of blaze angles to begin with (figure~\ref{fig:blazebend}, left).

\label{sect:bend}
\begin{figure} [ht]
\begin{center}
\includegraphics[width=0.7\textwidth]{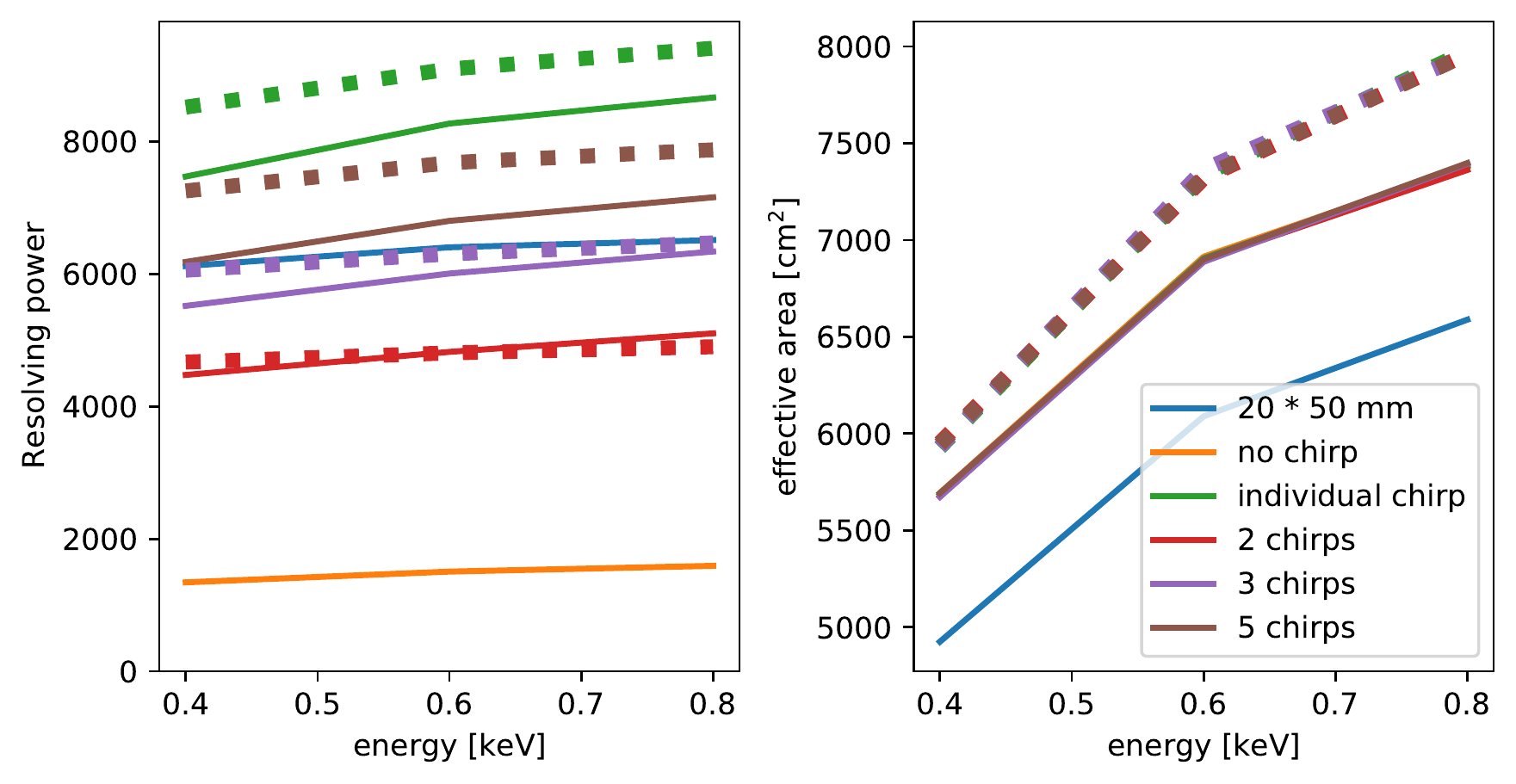}
\end{center}
\caption {\label{fig:chirp_aeff_res_bend}
Same as figure~\ref{fig:chirp_aeff_res}, but simulation for bent chirped gratings are added as dashed lines.
(The chirp is optimized for the bent gratings, i.e ``bend first, then chirp".)}
\end{figure}
In the previous section, we showed $R$ and $A_{\mathrm{eff}}$ for flat chirped gratings in figure~\ref{fig:chirp_aeff_res}. We now add simulations with bent chirped gratings and show the result in figure~\ref{fig:chirp_aeff_res_bend}. The resolving power is similar to or slightly better for bent chirped gratings than for flat chirped gratings, in particular if more than three different chirps are used. However, in practice, the alignment accuracy, which is not included in these simulations, also reduces the achievable resolving power. As long as $R > 5000$ in the figure, we can assume that other effects (alignment, pointing stability) limit the final value for $R$. 

Given that the gratings are located in a converging beam, for a flat grating, the blaze angle will change between the left and the right side of a grating. On the other hand, the focal length is large, so that difference is small, even for relatively large gratings. If not all orders are detected, then bent gratings might offer an advantage, because they follow the blaze angle description better and thus disperse more photons in the orders that are covered by detectors. For the three energies we ray-trace here, bending the gratings increases the effective area by 5-10\%.

\section{Summary and discussion}
\label{sect:summary}
The grating size for the DRM of the Lynx XGS is actually limited by the aberrations introduced by finite sized, flat gratings. A spectral resolving power $R$ of order 5000 can only be achieved with gratings no larger than about $20 * 50$~mm$^2$, if covering the full aperture, or $50 * 50$~mm$^2$ if sub-aperturing is used such that the areas of the aperture where the departure from the Rowland torus is largest remain uncovered. For Lynx, a few thousand gratings are needed to cover the aperture. We show here that chirped gratings can be used at much larger sizes, so that only a few hundred gratings are needed and the effective area is increased by about 25\% because less area is lost for grating frames and mounting structures. The optimal chirp is different for every grating position, but filling the aperture with gratings with just five different chirps still matches the resolving power requirement of Lynx. Given other non-ideal effects, such as mechanical alignment, using five different chirps is sufficient such that the accuracy of the chirp is no longer the limiting factor for the system resolving power.
We also investigated gratings that are bent and chirped, and found that bending large gratings can increase the effective area by another 5-10\%.

The results are relative numbers between scenarios with small flat gratings, large flat gratings with chirp, and large bend gratings with chirp. While we presented a scenario where the entire aperture is covered with gratings, and thus the effective area significantly exceeds the Lynx XGS science requirement, the relative change will hold up with sub-aperturing, i.e.\ we can increase the effective area to about 25\% compared to the DRM (where $2/3$ or the aperture is filled with gratings) when we fill the same space with large, chirped gratings, and by another 10\% when bending those gratings.

In Ref.~\citenum{CATXGS} we also discussed alignment tolerances. Using larger gratings impacts the alignment tolerances in some degrees of freedom. Any change that is global to all gratings (e.g.\ the position of the XGS relative to the detectors) still has the same alignment tolerances. Similarly, translations have the same effect, because we are just translating one large instead of several small gratings. What does change are the limits on the rotation of individual gratings. Because of the larger size, the gratings have a longer ``lever arm'', so a rotation of a grating around its own center will make the edges of a large grating deviate more from the design position than for a small grating. Thus, the tolerances for grating rotations around the dispersion and cross-dispersion scale inversely with the grating size.

In summary, we showed in this work that the effective area of the Lynx XGS can be increased by 25\% using large chirped gratings and another 10\% by bending these gratings, without detrimental impact on the resolving power.

\acknowledgments 
Support
for this work was provided in part through NASA grant NNX17AG43G and
Smithsonian Astrophysical Observatory (SAO) contract SV3-73016 to MIT
for support of the {\em Chandra} X-Ray Center (CXC), which is operated
by SAO for and on behalf of NASA under contract NAS8-03060.  The
simulations make use of Astropy, a community-developed core Python
package for Astronomy\cite{astropy1,astropy2}, numpy\cite{numpy}, and
IPython\cite{IPython}. Displays are done with mayavi\cite{mayavi} and
matplotlib\cite{matplotlib}.

\bibliography{report} 
\bibliographystyle{spiebib} 

\end{document}